\documentclass[amsmath,amssymb,superscriptaddress]{revtex4}
\input{psfig.sty}
\begin{document}
%
\title{Cluster meanfield approximation for the optical response of
weakly doped semiconductor quantum wells}
\author{Franz X.~Bronold}
\affiliation{Institut f\"ur Theoretische Physik,
Otto-von-Guericke-Universit\"at Magdeburg,
D-39016 Magdeburg,  Germany}
\affiliation{Physikalisches Institut, Universit\"a{}t Bayreuth,
D-95440 Bayreuth, Germany}
\date{\today}      

\begin{abstract}
The calculation of the optical properties of doped semiconductor
quantum wells 
is an intricate many-body problem because of the {\bf dynamical response 
of the excess carriers to the photogenerated valence band hole}.
At low densities, however, where the main effect of the dynamical 
response is the formation of {\bf trions}, a simple {\bf cluster 
meanfield approximation} can be effectively employed to calculate 
the {\bf optical susceptibility}. 
\end{abstract} 
\maketitle  

\section{Introduction}

The interaction of an optically generated exciton 
with excess carriers in the conduction band has been the subject 
of extensive experimental and theoretical investigations. 
Especially the high density regime attracted much attention and is 
qualitatively understood \cite{HS84,SCM89,Zimmermann88}: 
Dynamical screening and 
Pauli blocking prohibit the formation of bound $eh$~\cite{notation}
states, 
the exciton is therefore unstable and its spectral weight  
distributed among $eh$ scattering states. Residual 
correlations between the valence band (VB) hole and conduction band
(CB) electrons are still important,
however, and, depending on the CB electron to VB hole mass ratio,
give rise to a more or less pronounced Fermi edge singularity~\cite{Brum97},
for instance, in optical absorption spectra.  

With the seminal experiment of Kheng et al.~\cite{Kheng93},
which showed that in a weakly n-doped semiconductor quantum well
(QW), photogenerated $eh$ pairs give rise to 
trions, i.e., bound $eeh$ states
comprising two CB electrons and one VB hole, the center of interested shifted
to low densities. As a result of numerous theoretical and 
experimental studies~\cite{Usukura99,Ruan99,Stebe89,Stebe97,Stebe98,Bronold00,
Esser00,Esser01,Suris01,
Lovisa97,Siviniant99,Eyton98,Manassen96,Shields95,Brown96,Yusa00,trion},
the importance of $eeh$ states in the low density regime is now 
unambiguously established and 
a qualitative understanding of the low density regime is also starting 
to emerge. 

\begin{figure}[t]
\hspace{0.0cm}\psfig{figure=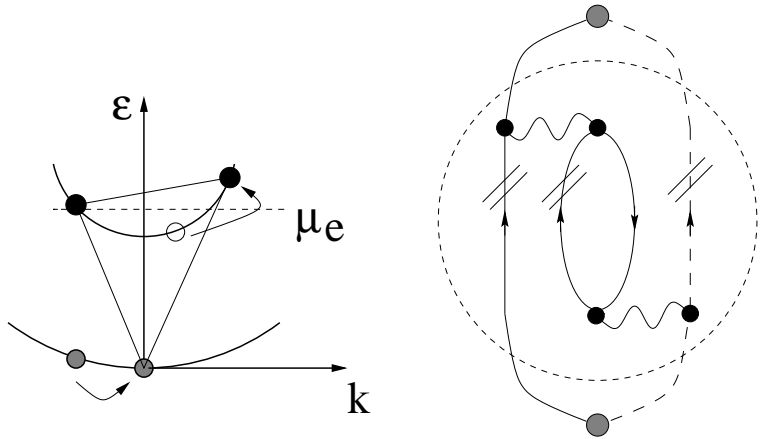,height=5.77cm,width=10.0cm,angle=0}
\caption[fig1]
{The left panel illustrates a typical dynamical response  
of the CB electrons to the photogenerated VB hole in an idealized 
n-doped semiconductor ($\mu_e$ is the chemical potential for the CB 
electrons). The right panel shows the corresponding Feynman diagram. The dashed 
forward running line denotes a VB hole, whereas the forward and backward 
running solid lines depict, respectively, a CB electron and CB hole.
Multiple scattering, indicated by the thin stripes attached to the 
participating lines,  
correlates the intermediate $eeh$ cluster to a trion.}
\label{fig1}
\end{figure}

Both the Fermi edge singularity and
the trions are consequences of the dynamical 
response of the CB electrons to the sudden appearance of the 
photogenerated VB hole. A typical lowest order process is depicted
in Fig. [\ref{fig1}]. The photogenerated VB hole scatters from one
momentum state to another and simultaneously excites a virtual 
$e\bar{e}$ pair (recall our notation~\cite{notation})
to compensate for the momentum transfer. At 
low densities, the Coulomb interaction is strong enough to 
correlate the VB hole with two CB 
electrons (the photoexcited CB electron and the CB electron from 
the $e\bar{e}$ pair) and a (negatively charged) trion 
appears. At higher CB electron densities, 
an {\it arbitrary} number of $e\bar{e}$ pairs is 
excited, giving rise to, among others, the screening of the 
Coulomb interaction. At some density, the Coulomb interaction is 
then too weak to support trions and/or excitons, but the residual
correlations might produce a Fermi edge singularity.  

The close connection between trion formation at low densities 
and the appearance of the Fermi edge singularity at high densities has 
been verified experimentaly~\cite{Shields95,Brown96,Yusa00} 
but, so far, a unified theoretical 
description of the full density dependence of the optical 
susceptibility, based, e.g., on the qualitative  
considerations of the previous paragraph, is still missing.
To construct such a theory is a rather formidable task. Here, we focus 
therefore on the low density regime, where the cluster meanfield 
approximation (CMFA)~\cite{Dukelsky98} can be effectively used to 
calculate the 
optical susceptibility.~\cite{Bronold00} In contrast to early
theoretical studies of trion 
states~\cite{Usukura99,Ruan99,Stebe89,Stebe97,Stebe98}, 
the CMFA describes trion and exciton states simultaneously. Moreover, it 
explicitly accounts for a (low) density of excess carriers. [In this 
sense it is closely related to the density matrix method of 
Esser et al.~\cite{Esser01} and the exciton-electron T-matrix 
model of Suris et al.~\cite{Suris01}.] 
The high density regime, however, cannot be systematically reached,
despite the earlier optimistic assessment of the author. 

\section{Cluster meanfield approximation (CMFA)}

\subsection{Optical susceptibility}

We consider for simplicity an idealized strictly two-dimensional 
quantum well with two isotropic parabolic bands. With an 
appropriate choice of single particle states, our formalism can 
be of course also applied to more realistic QW models, which 
include, e.g., valence band mixing and finite band off-sets.  
We assume a single hole in the VB and a low concentration $n$ 
of electrons in the CB. 
The charge carriers are coupled through the
Coulomb interaction and the total Hamiltonian is given by
\begin{eqnarray}
H&=&H_0 + V\nonumber\\
&=&\sum_{\vec{k}\sigma}\large[\epsilon_c(k)a_{\vec{k}\sigma}^\dagger 
a_{\vec{k}\sigma}
+\epsilon_v(k) b_{\vec{k}\sigma}^\dagger b_{\vec{k}\sigma}\large]
+{1\over 2}\sum_{\vec{k}\vec{p}\vec{q}\sigma\tau}v(q)
\large[a_{\vec{k}+\vec{q}\sigma}^\dagger a_{\vec{p}-\vec{q}\tau}^\dagger 
a_{\vec{p}\tau} a_{\vec{k}\sigma}
-2 a_{\vec{k}+\vec{q}\sigma}^\dagger b_{\vec{p}-\vec{q}\tau}^\dagger 
b_{\vec{p}\tau}a_{\vec{k}\sigma}\large],
\label{model}
\end{eqnarray}
where $\epsilon_c(k)=E_g+k^2/2m_e$ and
$\epsilon_v(k)=k^2/2m_h$ are the CB and VB dispersions ($\hbar=1$),
respectively, and $V(q)=4\pi e^2/\epsilon_0q$ is the
Coulomb interaction in standard notation. 

If the momentum dependence of the dipole matrix element
$r_\gamma$ is ignored, linear 
response theory gives for the optical susceptibility 
[`1' stands for $\vec{k}_1$]  
\begin{eqnarray}
\chi(\omega)=2|r_\gamma|^2\sum_{12}
\langle\langle b_{-1}a_{1};a_{2}^\dagger b_{-2}^\dagger\rangle\rangle_\omega
\label{chi1}
\end{eqnarray} 
where$\langle\langle b_{-1}a_{1};a_{2}^\dagger b_{-2}^\dagger\rangle\rangle_\omega$
denotes the Fourier transform of the  
two-time correlation function
\begin{eqnarray}
\langle\langle (b_{-1}a_{1})(t);(a_{2}^\dagger b_{-2}^\dagger)(0)\rangle\rangle
=-i\Theta(t)\langle [(b_{-1}a_{1})(t),(a_{2}^\dagger b_{-2}^\dagger)(0)] \rangle.
\label{Pt}
\end{eqnarray}
Here, the thermodynamic expectation value is defined by 
$\langle ... \rangle=Z^{-1}~Tr(e^{-\beta{\cal H}}...)$, 
with $Z=Tre^{-\beta{\cal H}}$, and ${\cal H}=H-\mu_e N_e$.
The imaginary part of $\chi(\omega)$ determines, apart from 
a constant factor, the optical absorption 
$I_{abs}(\omega)\sim -Im\chi(\omega)$. The overall factor of two in front 
of the sum on the rhs of Eq. (\ref{chi1}) comes from the spin.  

\subsection{Dyson-type equations}

The CMFA, originally developed to attack the 
nuclear many-body problem, rearranges the equations of
motion for two-time correlation functions into an hierarchy of Dyson-type
equations. It is thus closely related to the Mori memory 
function method, although the mathematical details 
are quite different. 
Using CMFA techniques~\cite{Dukelsky98}, 
we can derive a formally exact Dyson-type equation for the $eh$ pair 
propagator $P(12;\omega)=
\langle\langle b_{-1}a_{1};a_{2}^\dagger b_{-2}^\dagger\rangle\rangle_\omega$.
Keeping in mind that the spin configurations of the various correlation 
functions are 
fixed, because of the spin-independence of the Coulomb scattering, 
we suppress the spin variables and write:   
\begin{eqnarray}
[\omega+i\eta-Z(1)]P(12;\omega)=N(1)\{\delta_{12}
+\sum_3[M^{st}(13) + \delta M(13;\omega)]P(32;\omega),
\label{Dyson}
\end{eqnarray}
with
\begin{eqnarray}
M^{st}(12)&=&N^{-1}(1)N^{-1}(2)
\langle[[b_{-1}a_1,V],a_2^\dagger b_{-2}^\dagger]\rangle\\
\delta M(12;\omega)&=&N^{-1}(1)N^{-1}(2)
\langle\langle
[b_{-1}a_1,V];[V,a_2^\dagger b_{-2}^\dagger]\rangle\rangle_\omega\\   
N(1)&=&\langle [b_{-1}a_1,a_2^\dagger b_{-2}^\dagger]\rangle\\
Z(1)&=&\epsilon_c(1)+\epsilon_v(-1)
\end{eqnarray}
Applying the CMFA technique to the correlation function defining 
$\delta M(12;\omega)$, and so on, we obtain an hierarchy of 
Dyson-type equations for correlation functions with an increasing
number of particles. Clearly this set of equations has to be 
truncated and it is the truncation which restricts the CMFA 
to low CB electron densities.  

We now briefly recall the truncation procedure adopted in Ref. 
\cite{Bronold00}. The vanishing VB hole concentration implies 
$\langle b_1^\dagger b_1\rangle=0$, i.e., 
$N(1)=1-\langle a_1^\dagger a_1\rangle$. By assumption, the CB electron 
density is also small, i.e., $N(1)\simeq 1$. Thus, 
in leading order in the CB electron density, we 
ignore Pauli blocking. In the low density regime, we also neglect 
screening and single-particle selfenergy corrections
and Eq. (\ref{Dyson}) reduces to
\begin{eqnarray}
P(12;\omega)=
P^{(0)}(12;\omega)
+\sum_{34}P^{(0)}(13;\omega)
[M^{st}(34)+\delta M(34;\omega)]
P(42;\omega),
\label{Dysondilute}
\end{eqnarray}
with a bare $eh$ pair propagator, 
$P^{(0)}(12;\omega)=
\delta_{12}[\omega+i\eta-Z(1)]^{-1}$, 
and a static part of the $eh$ pair selfenergy $M^{st}(45)=-v(4-5)$.
The dynamical response of the CB electrons, i.e., the
creation of virtual $e\bar{e}$ pairs, is encoded in 
the dynamical (resonating) part of the $eh$ pair selfenergy, which, 
in the low density regime, becomes      
\begin{eqnarray}
\delta M(12;\omega)&=&\sum_{4567}v(5)v(7)
\Large(R(1,4,4+5,5-1|2,6,6+7,7-2;\omega)\nonumber\\ 
&-&R(1-5,4,4-5,-1|2,6,6+7,7-2;\omega)\nonumber\\
&-&R(1,4,4+5,5-1|2-7,6,6-7,-2;\omega)\nonumber\\
&+&R(1-5,4,4-5,-1|2-7,6,6-7,-2;\omega)\Large).
\label{dM}
\end{eqnarray}
We defined an eightpoint $ee\bar{e}h$ function 
$R(1235|5678;\omega)=\langle\langle b_4a_3^\dagger a_2a_1; 
a_5^\dagger a_6^\dagger a_7b_8^\dagger\rangle\rangle_\omega$, 
for which,as indicated above, we again derive a Dyson-type equation, 
symbolically written as 
$R(\omega)=R^{(0)}(\omega)+R^{(0)}(\omega)[K^{st}+\delta K(\omega)]R(\omega)$. 
[Assuming that the spin configuration of the photogenerated $eh$ pair 
is ($\uparrow\downarrow$), the spin configuration of the $ee\bar{e}h$ 
cluster defining $R$ is 
($\uparrow\sigma\sigma\downarrow$); $\sigma$ is then summed over 
in Eq. (\ref{dM}).] 
In the dilute limit, we need $\delta M(12;\omega)$ 
only to $o(n)$. Using an $\bar{e}$-line expansion for
$\delta M(12;\omega)$, we can show 
(i) that the term $\delta K(\omega)$ gives rise to   
$o(n^2)$ contributions to $\delta M(12;\omega)$ and is thus 
negligible at low densities, and (ii) that, in leading order 
in the CB electron density, 
the eightpoint $ee\bar{e}h$
function appearing in Eq. (\ref{dM}) in fact factorizes into
\begin{eqnarray}
R(1234|5678;\omega)=\delta_{37}f_e(3)G_3(12|56;\omega+\epsilon_c(3)),
\end{eqnarray}
with $f_e(3)=<a_3^\dagger a_3>$ and a sixpoint $eeh$ function 
$G_5(12|34;\omega)
=\langle\langle b_{5-1-2}a_2a_1;
a_3^\dagger a_4^\dagger b_{5-3-4}^\dagger\rangle\rangle_\omega$, which
satisfies  
\begin{eqnarray}
[\omega+i\eta-Z_5(12)]G_5(12|34;\omega)=I(12|34)
+\sum_{67}{\cal V}_5(12|67)G_5(67|34;\omega),
\label{sixpoint}
\end{eqnarray}
with 
\begin{eqnarray}
{\cal V}_5(12|67)=v(7-2)\delta_{6,1+2-7}
-v(1-6)\delta_{2,7}-v(2-7)\delta_{6,1},
\end{eqnarray}
and
\begin{eqnarray}
Z_5(12)&=&\epsilon_c(1)+\epsilon_c(2)+\epsilon_v(5-1-2).  
\end{eqnarray}
Thus, at low densities, the hierarchy of Dyson-type equations can 
be truncated at the level of a sixpoint $eeh$ function and,  
in leading order in the CB electron density, 
the optically generated $eh$ pair is only coupled
to an $eeh$ cluster. The groundstate of the $eeh$ cluster has
antiparallel CB electron spin. Focusing on the $eeh$ groundstate,
we fix therefore the spin configuration of the $eeh$ cluster defining 
$G$ to ($\uparrow\downarrow\downarrow$). For this spin configuration,
$I(12|34)=\delta_{13}\delta_{24}$,
and the spin eventually leads to the overall factor of two in 
Eq. (\ref{chi1}) for the optical susceptibility.

Using the spectral representation for the sixpoint $eeh$ function,
\begin{eqnarray}
G_5(12|34;\omega)=\sum_n 
{{\Psi_{5n}(12)\Psi_{5n}^*(34)}\over{\omega+i\eta-\Omega_n(5)}},
\end{eqnarray}
we finally obtain for the dynamic part of the $eh$ pair selfenergy
\begin{eqnarray}
\delta M(34;\omega)=
\sum_{5n} f_e(5)
{{\Xi_{5n}(3)\Xi_{5n}^*(4)}
\over{\omega+i\eta+\epsilon_c(5)-
\Omega_n(5)}},
\label{Self}
\end{eqnarray} 
where all multiple scattering events within the $eeh$ cluster
are now encoded in a vertex function
\begin{eqnarray}
\Xi_{5n}(3)=
\sum_{7} v(5-7)
\Large[\Psi_{5n}(37)-\Psi_{5n}(7,3+5-7)\Large],
\end{eqnarray}
given in terms of (normalized and complete) momentum-space $eeh$ wavefunctions
$\Psi_{5n}(37)$, which satisfy an $eeh$ momentum-space Schr\"odinger equation:
\begin{eqnarray}
[\Omega_n(5)-Z_5(12)]\Psi_{5n}(12)-\sum_{34}{\cal V}_5(12|34)\Psi_{5n}(34)=0.
\label{LS}
\end{eqnarray}
Here, `n' and `5' depict, respectively, internal quantum numbers
and the (center-of-mass) momentum of
the (propagating) bound $eeh$ cluster.  

To calculate $P(12;\omega)$ it is advantageous to split Eq. (\ref{Dysondilute}) 
into two coupled integral equations with, respectively, 
$M^{st}(=-v)$ and $\delta M$ as 
kernels: 
\begin{eqnarray}
P^{st}(12;\omega)&=&P^{(0)}(12;\omega)-
\sum_{34}P^{(0)}(13;\omega)v(3-4)P^{st}(42;\omega)
\label{Dysonstatic}\\
P(12;\omega)&=&P^{st}(12;\omega)+
\sum_{34}P^{st}(13;\omega)\delta M(34)P(42;\omega)
\label{Dysondynamic}
\end{eqnarray}
The first integral equation is readily solved using  
again a spectral representation 
\begin{eqnarray}
P^{st}(12;\omega)=\sum_\nu
{{\Phi_\nu(1)\Phi_\nu^*(2)}\over{\omega+i\eta-E_\nu}},
\end{eqnarray} 
where the (normalized and complete) momentum-space exciton wave 
functions $\Phi_\nu(1)$ obey an $eh$ momentum-space Schr\"odinger equation 
(Wannier equation)
\begin{eqnarray}
[E_\nu-Z(1)]\Phi_\nu(1)+
\sum_2 v(1-2)\Phi_\nu(2)=0.
\label{Wannier}
\end{eqnarray}
The solutions of this equation are analytically known.~\cite{Haug90}
  
\begin{figure}[t]
\hspace{0.0cm}\psfig{figure=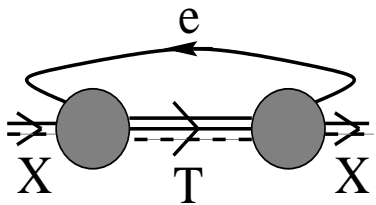,height=2.605cm,width=5.0cm,angle=-0}
\caption[fig2]
{Graphical illustration (not a Feyman diagram)
of the dynamical part of the exciton selfenergy 
$\delta M_{X}(\omega)$.
The full circles denote the exciton-trion coupling constants $u_{5T}^X$
and  $[u_{5T}^X]^*$.}
\label{fig2}
\end{figure}      

The second integral equation cannot be solved exactly. In order to  
avoid heavy numerics, we employ for its solution a
simple approximation scheme based on the ``exciton representation'', i.e., 
on a bilinear expansion  
in terms of the solutions of Eq.~(\ref{Wannier}). Writing
$P(12;\omega)=\sum_{\nu,\mu} P_{\nu\mu}(\omega)\Phi_\nu(1)^*\Phi_\mu(2)$,
we transform Eq. (\ref{Dysondynamic}) into an infinite 
set of algebraic equations for the expansion coefficients, 
\begin{eqnarray}
[\omega+i\eta-E_\nu]P_{\nu\mu}(\omega)=\delta_{\nu\mu}
+\sum_{\lambda}\delta M_{\nu\lambda}(\omega)P_{\lambda\mu}(\omega),
\label{Dysonalgebra}
\end{eqnarray}
where 
\begin{eqnarray}
\delta M_{\nu\lambda}(\omega)=
\sum_{5n}f_e(5){{u_{5n}^\nu[u_{5n}^\lambda]^*}\over
{\omega+i\eta+\epsilon_c(5)-\Omega_n(5)}},
\label{Mdyn}
\end{eqnarray}
and $u_{5n}^\nu=\sum_1\Phi_\nu^*(1)\Xi_{5n}(1)$.
In the vicinity of the exciton resonance, i.e., for energies
$\omega$ close to the exciton groundstate energy $E_X$, we 
expect the exciton and trion groundstate to be dominant. Thus, we 
solve Eq. (\ref{Dysonalgebra}) by employing one-pole
approximations: 
First, we keep only the term 
corresponding to the exciton ground state $\nu=X$ and 
neglect the coupling to excited exciton states $\nu\neq X$.
Next, assuming that the main modification of the 
exciton ground state comes from the groundstate of the $eeh$ cluster,
i.e., from the trion denoted by $n=T$, we restrict 
the sum in Eq. (\ref{Mdyn}) to the term $n=T$. 

Let us emphasize, the one-pole approximations are by no means 
mandatory 
and various improvements are possible. 
For example, if the 
exciton binding energy is small, i.e., if  
the exciton groundstate is not well separated from the
first excited state, the first excited state 
has to be kept. This gives rise to two 
coupled algebraic equations (two-pole approximation).  
It is even conceivable
to treat all $eh$ scattering states in a similar way, introducing 
an ``effective'' excited state, which carries the spectral 
weight of the whole continuum. Excited trion
states can be also taken into account by simply extending the 
sum in Eq. (\ref{Mdyn})to the respective states. 

The theoretical work by Esser et al.~\cite{Esser01} indicates that these 
refinements are indeed necessary to obtain 
good quantitative agreement between theory and experiment. 
For the purpose of demonstration, however, we neglect here 
scattering states and stick to the simple one-pole approximation
described above. The optical susceptibility in the vicinity of 
the exciton resonance is then given by 
\begin{eqnarray}
\chi(\omega)\simeq {{2|s_X|^2}\over{\omega+i\eta-E_X-\delta M_X(\omega)}},
\end{eqnarray}
with an exciton selfenergy (see Fig. \ref{fig2})
\begin{eqnarray}
\delta M_X(\omega)=
\sum_{5}f_e(5){{|u_{5T}^X|^2}\over
{\omega+i\eta+\epsilon_c(5)-\Omega_T(5)}},
\label{Xself1}
\end{eqnarray} 
and an exciton oscillator strength $s_X=r_\gamma\sum_1\Phi_X(1)$. 
Note, the one-pole approximation
is conceptually close to the T-matrix model considered
by Suris et al. in their analysis of
absorption and reflection spectra of modulation-doped
quantum wells.~\cite{Suris01}             

To proceed further, we assume that the CB electron density is 
such that the Fermi function in Eq. (\ref{Xself1}) can be 
replaced by a Boltzmann function parametrized by the CB 
electron density $n$. 
Measuring energies in units of the 2D exciton
binding energy $4R$ ($R$ is the 3D exciton Rydberg) and length
in units of the 3D Bohr radius $a_B$, we define 
$\tilde{\chi}(\tilde{\omega})\equiv\chi(4R\tilde{\omega}+E_g)/4R$, with
$\tilde{\omega}=[\omega-E_g]/4R$, and rewrite the optical susceptibility 
into dimensionless form 
\begin{eqnarray}
\tilde{\chi}(\tilde{\omega})={{|s_X|^2}\over{8R^2}}
{1\over {\tilde{\omega}+i\tilde{\eta}+1-\delta\tilde{M}_X(\tilde{\omega})}}.
\end{eqnarray}
Here, $\tilde{\eta}=\eta/4R$ and 
\begin{eqnarray}
\delta\tilde{M}_X(\tilde{\omega})&\equiv&\delta M_X(4R\tilde{\omega}+E_g)/4R
\\
&=&\tilde{\beta}{M_T\over M_X}(na_B^2)
|u_T^X|^2 I(\tilde{\omega}),       
\label{Xself2}
\end{eqnarray}
where $\tilde{\beta}=4\beta R$, $M_X$, and $M_T$ are,
respectively, the thermal energy measured in
units of $4R$, the exciton mass, and the trion mass.
The exponential integral, 
\begin{eqnarray}
I(\tilde{\omega})=
\int_0^\infty dy
{{e^{-\tilde{\beta}{M_T\over M_X}y}}\over{\tilde{\omega}+
i\tilde{\eta}-\tilde{\epsilon}_T+y}},
\label{ExpInt}
\end{eqnarray}
where $\tilde{\epsilon}_T=[\Omega_T(0)-2E_g]/4R$ 
is the energy
of a trion at rest, i.e., the
{\it internal} trion energy, which is the binding energy of {\it two}
CB electrons to one VB hole. 
The dimensionless exciton-trion coupling constant $u_T^X$ finally is  
given by an integral involving the wave function of 
the trion groundstate, the Coulomb interaction, and the 
wavefunction of the exciton groundstate. For simplicity we neglected 
the weak dependence of $u_T^X$ on the (center-of-mass) 
momentum of the trion. Note, for vanishing CB electron density $n$, 
$\delta\tilde{M}_X(\tilde{\omega})\rightarrow 0$ and 
$\tilde{\chi}(\tilde{\omega})$ describes a single exciton resonance.  

To obtain the exciton-trion coupling constant $u_T^X$ and the
internal trion energy $\tilde{\epsilon}_T$,
we employ the variational technique originally used by 
St\'eb\'e and coworkers~\cite{Stebe89,Stebe97} to calculate various 
properties of an isolated trion. 
For that purpose we transform 
Eq. (\ref{LS}) to real space (to solve the $eeh$ problem we use atomic 
units, which are, however, at the end of the calculation, translated 
into the excitonic units adopted above)
\begin{eqnarray}
[T+V]\Psi_T(\vec{r}_{e1}\vec{r}_{e2}\vec{r}_{h})=
\epsilon_T\Psi_T(\vec{r}_{e1}\vec{r}_{e2}\vec{r}_{h}),
\nonumber
\end{eqnarray}
with kinetic and potential energies ($\hbar=1$),
\begin{eqnarray}
T&=&-{1\over{2m_e}}(\Delta_{e1}+\Delta_{e2})-{1\over{2m_h}}\Delta_{h},
\nonumber\\
V&=&-{e^2\over{\epsilon_0}}
({ 1 \over {|\vec{r}_{e1}-\vec{r}_h|} }
+{ 1 \over {|\vec{r}_{e2}-\vec{r}_h|} }
-{ 1 \over {|\vec{r}_{e1}-\vec{r}_{e2}|} }),
\nonumber
\end{eqnarray}   
separate away the center-of-mass motion, which is simply a plane wave, 
and make a variational Hylleraas Ansatz for the internal part 
of the trion groundstate 
wavefunction,  
\begin{eqnarray}
\Psi(s,t,u)=\sum_{lnm} c_{lnm} |lnm\rangle,
\label{Hylleraas}
\end{eqnarray}
where $|lnm\rangle=e^{-{s\over 2}} s^l t^m u^n$ and 
$u, s$, and $t$ are ellipitical coordinates for the 
internal motion of the $eeh$ cluster. Note, since the trion
groundstate is a singlet, $\Psi(s,t,u)$ has to be an even 
function in $t$, i.e., $m$ has to be even. The expansion
coefficients are determined by a Ritz variational 
principle, i.e., by minimizing the energy functional
$E[\vec{c},\epsilon_T]=\langle\Psi|T+V|\Psi\rangle/\langle\Psi|\Psi\rangle$.
The variation yields an eigenvalue 
problem which is then iteratively solved. The largest 
eigenvalue is related to $\epsilon_T$ ($\rightarrow \tilde{\epsilon}_T$)
and the associated 
eigenfunction gives the expansion coefficients $c_{lnm}$. 
For more details we refer the readers to Refs.~\cite{Stebe89,Stebe97} 
and~\cite{BetheSalpeter}.
With the Hylleraas variational function, 
the exciton-trion coupling constant becomes 
\begin{eqnarray}
u_T^X=\sqrt{{2\over{N\pi}}}{1\over{k(1+\sigma)}}\int_0^\infty ds\int_0^s du\int_0^u dt
{{2\pi(s^2-t^2)u}\over{\sqrt{(u^2-t^2)(s^2-u^2)}}}
e^{-{{s+t}\over{k(1+\sigma)}}}
[{2\over{s-t}}-{1\over u}]\Psi(s,t,u),
\end{eqnarray}   
where $k$ is the effective charge of a CB electron,  
$N$ is the norm of $\Psi$, and $\sigma=m_e/m_h$.

\subsection{Some numerical results}

Using a 2D 22-term Hylleraas wavefunction, we calculated in Ref.~\cite{Bronold00} 
the optical absorption along the
lines explained in the previous subsections. 
For a mass ratio $\sigma=m_e/m_h=0.146$,
corresponding to GaAs, we found 
$u_T^X\simeq 1.0896$ and 
$\tilde{\epsilon}_T=-1.1151$. ($u_T^X$ was obtained by Gaussian 
quadrature.) The binding energy (in units of 
$4R$) for {\it one} electron is therefore $\tilde{W}_T=0.1151$ 
in agreement with the binding energies obtained by other 
means.~\cite{Usukura99} 

\begin{figure}[t]
\hspace{0.0cm}\psfig{figure=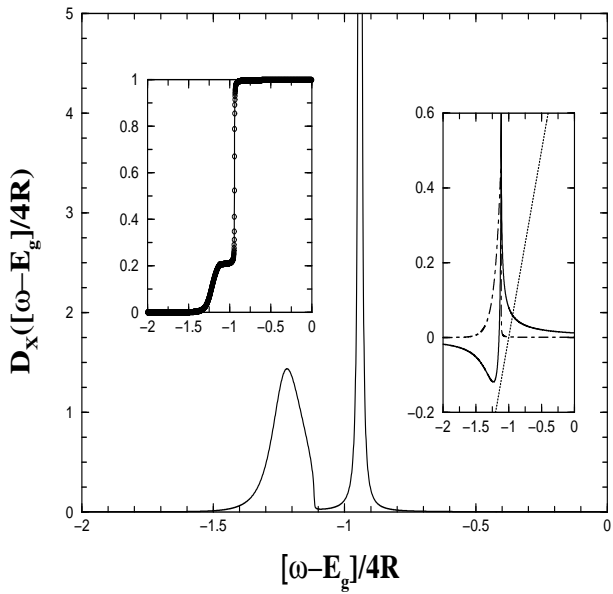,height=8.0cm,width=8.0cm,angle=-90}
\caption[fig3]
{The spectral function $D_X(\tilde{\omega})$ for
$na_B^2=0.008$ and $4R\beta=10$.
The effective mass ratio
$\sigma=0.146$, corresponding to GaAs, and
$\tilde{\eta}=0.001$.
The left inset depicts the integrated spectral weight
$C_X(\tilde{\omega})$,
whereas the right inset
displays $\tilde{\Gamma}_X(\tilde{\omega})$
(dot-dashed line) and
$\tilde{\Delta}_X(\tilde{\omega})$ (solid line). The energy
for which $\tilde{\Delta}_X(\tilde{\omega})$ crosses
$\tilde{\omega}+1$ (dotted line) defines the
exciton line at $\tilde{\omega}=-0.94$. From Ref.~\cite{Bronold00}.
}                          
\label{fig3}
\end{figure}

The optical absorption in the vicinity of the exciton resonance 
is related to the exciton spectral function 
\begin{eqnarray}
D_X(\tilde{\omega})=-Im\tilde{\chi}(\tilde{\omega}),
\end{eqnarray}
shown  in Fig. \ref{fig3} for $na_B^2=0.008$ and $\tilde{\beta}=10$. 
The overall structure is in qualitative
agreement with experiments, despite the simplicity of the model. 
Below a narrow 
peak at $\tilde{\omega}= -0.94$, we see a broad absorption band 
with a sharp high energy edge at $\tilde{\omega}=-1.1151$ and 
a low energy tail. The narrow peak is the exciton resonance whereas
the broad band corresponds to the absorption due to 
a trion. As shown in Ref.~\cite{Bronold00}, the sharp high energy edge 
of the absorption band comes from the pole of the exciton selfenergy at 
$\tilde{\epsilon}_T=-1.1151$, i.e., a photon in resonance with the 
high energy edge creates a trion with momentum zero.
The band has a tail at the low energy side because
of the recoil of the trion, i.e., the tail comprises trion states
with finite momentum. The right inset of Fig. \ref{fig3}
displays the real and imaginary parts of the exciton selfenergy,  
$\delta\tilde{M}_X(\tilde{\omega})=
\tilde{\Delta}_X(\tilde{\omega})-i\tilde{\Gamma}_X(\tilde{\omega})$. 
As can be seen, trion states with small momenta (high energy edge) are 
strongly damped. The maximum of the absorption band does therefore 
not correspond to a trion with zero momentum 
(as in the case of vanishing CB electron concentration~\cite{Stebe98})
but to a trion state with a finite momentum. 
The left inset depicts the integrated spectral weight up to an energy
$\tilde{\omega}$ defined in terms of the cumulant 
$C_X(\tilde{\omega})=\int_{-\infty}^{\tilde{\omega}} d\tilde{E}
D_X(\tilde{E})$. As a consequence of the one-pole approximation,  
the total spectral weight associated with the 
trion band and the exciton line adds up to one. If 
scattering states had been taken into account, they would 
also carry some spectral weight, i.e., in a complete theory, 
the combined spectral weight of trion and exciton has to be of course
less then one.  
Our exploratory calculation neglected scattering states,
because, at low enough CB densities, in particular, within the range of 
validity of our theory, scattering states are expected to carry almost no
spectral weight. Indeed, the more refined calculations of Esser  
et al.~\cite{Esser01} seem to verify this assessment. The spectral
weight, which can be associated with scattering states, is less 
then 5$\%$ of the 
spectral weight of the exciton groundstate.  

\section{Conclusions}

We presented a detailed description of the CMFA 
for the optical  
response of a weakly n-doped idealized semiconductor QW. 
The CMFA, tailormade
for low CB electron densities, reduces the calculation of the 
optical susceptibility to the solution of the Schr\"odinger 
equations for, respectively, a single $eh$ pair and a single 
$eeh$ cluster. Excess carriers are only treated as
a reservoir for bound state formation, which does not, in leading order
in the CB electron density, modify the
properties of the $eh$ and $eeh$ states. We derived the CMFA for an
idealized QW model, but it is straightforward to generalize the formalism 
to more realistic models. Furthermore, relaxing the one-pole approximations, 
which we solely used for illustration, would allow to take exciton and 
trion continuum states into account. 
At intermediate-to-high
CB electron density, many-body medium corrections, such as 
Pauli blocking, screening, and single particle selfenergy
corrections, substantially modify $eh$ and $eeh$ states.   
Within the CMFA it is however not clear how to systematically account
for these effects. 



\begin{references}
\bibitem{notation}
$e$, $\bar{e}$, and $h$ denote, respectively, CB electrons, CB holes,
and VB holes; a trion, e.g., is therefore a $eeh$ cluster.
\end{references}
\end{document}